%% file: hello_sme.tex
\documentclass[conference]{IEEEtran}
\usepackage{cite}
\usepackage{amsmath,amssymb,amsfonts}
\usepackage{algorithmic}
\usepackage{graphicx}
\usepackage{textcomp}
\usepackage{xcolor}
\usepackage{multirow}
\usepackage{caption}
\usepackage{subcaption}
\PassOptionsToPackage{hyphens}{url}\usepackage{hyperref}
\usepackage{listings}
\makeatletter
\def\lst@PlaceNumber{%
  \makebox[3em][l]{\normalfont\lst@numberstyle{\thelstnumber}}%
}
\makeatother

\begin{document}

\title{Hello SME!\\ Generating Fast Matrix Multiplication Kernels Using the Scalable Matrix Extension
}

\author{\IEEEauthorblockN{Stefan Remke}
\IEEEauthorblockA{\textit{Faculty of Mathematics and Computer Science} \\
\textit{Friedrich Schiller University Jena}\\
Jena, Germany \\
stefan.remke@uni-jena.de}
\and
\IEEEauthorblockN{Alexander Breuer}
\IEEEauthorblockA{\textit{Faculty of Mathematics and Computer Science} \\
\textit{Friedrich Schiller University Jena}\\
Jena, Germany \\
alex.breuer@uni-jena.de}
}

\maketitle

\begin{abstract}
  Modern central processing units (CPUs) feature single-instruction, multiple-data pipelines to accelerate compute-intensive floating-point and fixed-point workloads.
  Traditionally, these pipelines and corresponding instruction set architectures (ISAs) were designed for vector parallelism.
  In recent years, major hardware vendors have further increased the throughput of their CPUs by introducing matrix units with corresponding ISA extensions.
  The Scalable Matrix Extension (SME) has been announced for the Arm architecture in 2021 and Apple's M4 chip is the first to support SME.

  This paper presents an in-depth study of SME on M4.
  Our microbenchmarks determine the maximum floating-point and fixed-point throughput of M4's SME acceleration and study the achievable bandwidth for transfers to and from the matrix registers.
  Furthermore, we used the insights gained to design a just-in-time code generator for SME-based small matrix multiplications.

  The results presented show that M4's SME support is FP32-centric, with an achievable throughput of over 2.3\,FP32 TFLOPS.
  To maximize read and write bandwidth, loading and storing to and from the matrix registers must be done in two steps.
  Our just-in-time generated small matrix multiplication kernels outperform the vendor-optimized BLAS implementation in almost all tested configurations.
\end{abstract}


\input{sections/intro.tex}
\input{sections/sme.tex}
\input{sections/micro.tex}
\input{sections/jit.tex}
\input{sections/outlook.tex}

\bibliographystyle{IEEEtran}
\bibliography{references}
\end{document}

%% file: sections/intro.tex
\section{Introduction}
\label{sec:intro}
In 2008, Roadrunner, one of the first accelerated supercomputers, became operational \cite{2008_roadrunner}.
Since then, many heterogeneous systems with different types of accelerators have been built and successfully deployed for large-scale computing.
Prominent accelerated systems include Titan (NVIDIA K20X, 2012) \cite{2012_titan}, Tianhe-2 (Intel Knights Corner, 2013) \cite{2014_tianhe2}, Summit (NVIDIA V100, 2018) \cite{2019_summit}, Frontier (AMD MI250X, 2021) \cite{2023_frontier}, and Aurora (Intel Ponte Vecchio, 2023) \cite{2023_aurora}.

A traditional accelerated machine consists of a host CPU and one or more accelerator cards connected to the motherboard by an appropriate interconnect.
In this type of system, the CPU and the accelerator(s) have different address spaces.
This means that communication between the host and the device is explicit.
``Offloading'', where certain computationally intensive parts of the workload are offloaded to the accelerator(s), is the dominant programming model on such machines.

In recent years, the integration of CPUs and accelerators has become much tighter.
Especially in the smartphone market, system-on-a-chip (SoC) designs that tightly couple multiple different accelerators on a single chip have become the norm.
Well-known examples include Apple's A-series SoCs used in the iPhone \cite{2023_apple_a17}, Qualcomm's Snapdragon SoCs \cite{2023_qualcomm_snapdragon_8_gen3} and MediaTek's Dimensity-series \cite{2024_mediatek_dimensity7300_press}.
The introduction of Apple's M-series \cite{2024_apple_m4_press}, and products like Qualcomm Snapdragon X Elite \cite{2023_qualcomm_snapdragon_x_elite}, Intel Meteor Lake \cite{2022_meteor_lake} and AMD Phoenix/Hawk Point \cite{2024_8000g} also accelerated the use of SoCs in notebook and desktop computing.
In addition, SoCs targeting the HPC market have been introduced or announced.
Recent examples include AMD's Instinct MI300A \cite{2024_mi300a} and Intel's XPU strategy.
Accelerators on SoCs are typically used through tailored programming models.
For example, one might use HIP to target an integrated AMD GPU, Metal for an Apple GPU, or remote procedure calls to execute code on Qualcomm's VLIW Hexagon NPU. 
A key advantage of an SoC is that it typically allows us to access the same memory from both the CPU and the accelerator.

An even tighter coupling exists between recent matrix units and their host CPU cores.
Similar to floating-point coprocessors, these matrix units can be used directly through the CPU instruction stream via Instruction Set Architecture (ISA) extensions.
Typically, data can be copied from vector registers into a set of separate matrix registers, or loaded directly from caches into the matrix registers.
IBM introduced a matrix unit with Power10 (2021), resulting in Matrix-Multiply Assist (MMA) instructions \cite{2021_ibm_power10}.
Intel introduced matrix units in its Xeon processors with Sapphire Rapids (2023) \cite{2022_sapphire_rapids}, which supports Intel AMX.
In contrast, Apple has not released documentation for its proprietary ISA extension required to use matrix math acceleration in M1 (2020), M2 (2022), and M3 (2023).
However, the extension is known as ``Apple AMX'', and a community-driven effort at least partially documents its use \cite{2022_sme_eval}.

In 2021 Arm announced the Scalable Matrix Extension (SME) \cite{2021_sme_blog}, which is similar to Apple AMX at its core, but also introduces many features that are not documented for Apple AMX.
On May 7, 2024, Apple announced the M4 chip at its “Let Loose” event \cite{2024_apple_m4_press}.
M4 is the first publicly available silicon that supports SME.
This was confirmed by the community with microbenchmarks a few days after the SoC became available.
Later, on June 13, 2024, a pull request to the LLVM monorepository by Apple specifies ARMv9.2a for M4's CPU and confirms that it supports SME and SME2 \cite{2024_llvm_pr}.
This paper provides a thorough analysis of SME(2) performance on M4.
We make the following contributions:
\begin{itemize}
  \item Extensive benchmarking of M4's performance cores, efficiency cores, matrix acceleration, and memory subsystem.
  \item Implementation of a just-in-time SME(2) code generator for small matrix-matrix multiplications.
  \item Detailed performance analysis of the code generator and comparison with the vendor-optimized BLAS implementation in Accelerate.
\end{itemize}

%% file: sections/sme.tex
\section{Background}
This section provides a brief introduction to the used testbed in Sec.\,\ref{sec:testbed} and the studied ISA extension in Sec.\,\ref{sec:sme}.

\subsection{Testbed}
\label{sec:testbed}
All tests described in this paper were performed on a 2024 11-inch iPad Pro Wi-Fi 1 TB.
The tablet is equipped with an M4 chip with four performance cores and six efficiency cores, as well as 16 GiB of memory.
M4 is built in 3-nanometer technology and consists of 28 billion transistors \cite{2024_apple_m4_press}.
Both the performance and efficiency cores can issue Scalable Matrix Extension instructions.
The iPad runs iOS version 17.5 and was connected to a 2020 Mac mini via USB-C for all testing.

\subsection{Scalable Matrix Extension}
\label{sec:sme}
The Scalable Matrix Extension (SME) was added to the Arm Architecture Reference Manual for A-profile architecture on March 20, 2024.
SME defines the following key features \cite{2024_arm_manual}:
\begin{itemize}
    \item Architectural state capable of holding two-dimensional matrix tiles.
    \item A Streaming SVE processing mode.
    \item Instructions that accumulate the outer product of vectors into a tile.
    \item Load, store, and move instructions that transfer a vector to or from a tile row or column.
\end{itemize}
In addition, SME2 is a superset of SME and implements, among other features, data processing instructions that operate on groups of scalable vector registers.

We can enable access to the Streaming SVE (SSVE) mode and the SME architectural state by issuing the \lstinline{SMSTART} instruction.
Similarly, we can disable access to SSVE and SME by issuing the \lstinline{SMSTOP} instruction.
Most of the SSVE and SME data processing instructions operate on the scalable vector registers and the ZA array.
The respective sizes are determined by the Streaming Vector Length (SVL), which is 512 bits on Apple's M4.
SSVE defines the 32 vector registers \lstinline{Z0}-\lstinline{Z31} of length SVL.
The size of the ZA array in bytes is given by $\text{SVL}/8 \times \text{SVL}/8$ which is 4096\,bytes on M4.
Instructions access the ZA array either as vectors of SVL\,bits or through two-dimensional tiles, which are subarrays within ZA.

Outer product data processing instructions form the core of SME.
These instructions take two scalable vector registers as inputs and accumulate in a \lstinline{ZA} tile.
Outer product instruction mnemonics end with \lstinline{MOPA}.
An example is the floating-point outer product and accumulate instructions (\lstinline{FMOPA}).
In FP32 arithmetic, the FMOPA instruction computes the outer product of two $\text{SVL}/32$-element vectors located in two \lstinline{Z} vector registers.
The result is added to the data in a ZA tile with $\text{SVL}/32 \times \text{SVL}/32$ elements.
Since M4 has an SVL of 512\,bits, this effectively means that an FP32 \lstinline{FMOPA} instruction computes the outer product of two 16-element source vectors and adds the result to a \lstinline{ZA} tile with $16\times16$ elements.
Thus, a single FP32 \lstinline{FMOPA} instruction on M4 performs $16 \times 16 \times 2 = 512$ floating point operations (see also Fig.\,\ref{fig:fmla_fmopa}).

%% file: sections/micro.tex
\section{Microbenchmarks}
\label{sec:micro}
This section presents a collection of microbenchmarks that determine the performance characteristics of the heterogeneous CPU and associated matrix acceleration in Apple's M4 chip.
Sec.\,\ref{sec:bench_app} describes the overall structure of our iOS application used for benchmarking.
Next, in Sec.\,\ref{sec:micro_asimdneon} we study the floating-point performance when using ``traditional'' vector instructions, i.e., ASIMD/Neon instructions.
Then, in sections \ref{sec:micro_sme_outer}, \ref{sec:micro_sme_matrix} and \ref{sec:micro_sme_vector}, we make use of the Scalable Matrix Extension (SME/SME2) and examine the performance of the available outer product, matrix-matrix and vector instructions.
Finally, in Sec.\,\ref{sec:micro_multi_core} we study the multi-core performance of M4 and present bandwidth benchmarks in Sec.\,\ref{sec:micro_bandwidth}.

\subsection{Benchmarking App}
\label{sec:bench_app}
We have written all the microbenchmarks presented here in assembly language and call them from a C\texttt{++} wrapper.
In turn, we call the C\texttt{++} wrapper from a simple Swift application using another Objectiv-C wrapper.
The C\texttt{++} wrapper measures the elapsed time to execute a microbenchmark and handles multithreading.
We use Apple's Dispatch framework for multithreading.
Typical Linux-based pinning methods, such as using \lstinline{taskset}, are not supported on iOS.
So we use the available quality of service identifiers for concurrent queues in Dispatch.
In particular, we use \lstinline{QOS_CLASS_USER_INTERACTIVE} and \lstinline{QOS_CLASS_UTILITY}.
While Dispatch does not provide direct control over the executing cores, we have found that the performance of a thread in the user-interactive queue matches our expectations for a performance core, and the performance of a utility thread matches our expectations for an efficiency core.
When spawning multiple utility threads, we observe a maximum CPU utilization of 600\% in Xcode's monitoring tool, even when using more than six threads,
This indicates that the six efficiency cores are being used exclusively for utility threads.
In our tests, spawning more than four user-interactive threads results in CPU utilization of over 400\% indicating that user-interactive threads may also be running on the efficiency cores.

\subsection{Neon: Traditional Vector Instructions}
\label{sec:micro_asimdneon}
Our first set of benchmarks, studies the performance of vector instructions using Arm's ASIMD extension also called Neon.
The general kernel structure for determining the maximum FP32 performance of a core using Neon is given in Lst.\,\ref{lst:neon_fmla}.

The kernel takes the number of repetitions as a 64-bit input.
This parameter is passed through the general purpose register X0.
The body of the repeat loop (lines 4-8) consists of 30 independent FMLA (vector) instructions.
Neon specifies the 32 vector registers V0-V31, each of which is 128 bits wide.
FMLA (vector) multiplies the values in two source registers, adds the product to the values in the destination register, and writes the result back to the destination register.
For example, the instruction \lstinline{fmla v1.s, v30.s, v31.s} (line 5) multiplies the four FP32 values in V30 by those in V31, adds the product to the four FP32 values in V1 and writes the result to V1.
\begin{lstlisting}[caption={Assembly kernel for determining the maximum Neon floating-point throughput using the FMLA (vector) instruction and FP32 arithmetic.},label={lst:neon_fmla}]
// boilerplate code
repeat_loop:
  sub x0, x0, #1
  fmla v0.s, v30.s, v31.s
  fmla v1.s, v30.s, v31.s
  // 26 additional fmla instructions
  fmla v28.s, v30.s, v31.s
  fmla v29.s, v30.s, v31.s
  cbnz x0, repeat_loop
// boilerplate code
mov x0, #30*8
ret
\end{lstlisting}
In the kernel, we always use V30 and V31 as source registers and the remaining registers as destination registers to maximize the distance of occurring read-after-write dependencies on the destination registers.
Also, in each loop iteration, the number of outstanding repetitions is decremented by one (line 3).
The CBNZ instruction in line 9 jumps back to the label \lstinline{repeat_loop} (line 2) if there are any repetitions left.
The instruction in line 11 writes the number of floating-point operations performed in a single iteration to register X0, which is the kernel's return value.
Our C++ wrapper calls the kernel with a high enough number of repetitions so that the execution takes at least one second.

\begin{center}
  \begin{table}[b]
  \centering
  \begin{tabular}{|c|c|c|c|c|}
    \hline
    \multirow{2}{*}{\textbf{Instruction}} & \multicolumn{2}{c|}{\textbf{Datatype}} & \multicolumn{2}{c|}{\textbf{GOPS}} \\ [0.5ex]
    & \textbf{In} & \textbf{Out} & \textbf{P-Core} & \textbf{E-Core} \\ [0.05ex]
    \hline
    FMLA (Neon)   & FP64 & FP64 &   56 &  23 \\ [0.05ex]
    FMLA (Neon)   & FP32 & FP32 &  113 &  46 \\ [0.05ex]
    FMLA (Neon)   & FP16 & FP16 &  220 &  91 \\ [0.05ex]
    BFMMLA (Neon) & BF16 & FP32 &   67 &  31 \\ [0.05ex]
    \hline
    FMOPA (SME)   & FP64 & FP64 &  503 &  89 \\ [0.05ex]
    FMOPA (SME)   & FP32 & FP32 & 2009 & 357 \\ [0.05ex]
    \hline
    BFMOPA (SME)  & BF16 & FP32 & 2010 & 357 \\ [0.05ex]
    FMOPA (SME)   & FP16 & FP32 & 2010 & 357 \\ [0.05ex]
    SMOPA (SME)   &  I16 &  I32 & 2010 & 357 \\ [0.05ex]
    SMOPA (SME)   &   I8 &  I32 & 4017 & 715 \\ [0.05ex]
    \hline  
    FMLA (SME2)   & FP64 & FP64 & 251 &  89 \\ [0.05ex]
    FMLA (SSVE)   & FP64 & FP64 &  16 &  11 \\ [0.05ex]
    FMLA (SME2)   & FP32 & FP32 & 501 & 179 \\ [0.05ex]
    FMLA (SSVE)   & FP32 & FP32 &  31 &  22 \\ [0.05ex]
    \hline
  \end{tabular}
  \caption{Apple M4 performance when running different Neon, SSVE and SME(2) instructions on a single performance (P) or a single efficiency (E) core.}
  \label{tab:micro_single_core}
  \end{table}
\end{center}

We measured a Neon performance of 113 FP32 GFLOPS using a single performance core.
A single efficiency core has a performance of 46 FP32 GFLOPS when running the kernel in Lst.\,\ref{lst:neon_fmla}.
We also tested the performance of FP16 and FP64 FMLA (vector) instructions.
As shown in Tab.\,\ref{tab:micro_single_core}, the instruction throughput is the same, meaning that compared to FP32 arithmetic, we can perform twice as many FP16 operations per second and half as many FP64 operations per second.
The Neon BF16 matrix-matrix instruction BFMMLA has low floating-point throughput, making it an imperfect candidate for upstream kernels.

\subsection{Scalable Matrix Extension: Outer Product Instructions}
\label{sec:micro_sme_outer}
We use the FMOPA (non-widening) instruction to benchmark the FP32 performance of M4's matrix acceleration.
As discussed in Sec.\,\ref{sec:sme}, FMOPA computes the outer product of two vectors and adds the product to a matrix tile.
The corresponding assembly kernel is outlined in Lst.\,\ref{lst:sme_fmopa}.
\begin{lstlisting}[caption={Assembly kernel for determining the maximum SME floating-point throughput using the FMOPA (non-widening) instruction and FP32 arithmetic.},label={lst:sme_fmopa}]
// boilerplate code
ptrue p0.b
ptrue p1.b
repeat_loop:
  sub x0, x0, #1
  fmopa za0.s, p0/m, p1/m, z0.s, z1.s
  fmopa za1.s, p0/m, p1/m, z2.s, z3.s
  fmopa za2.s, p0/m, p1/m, z4.s, z5.s
  fmopa za3.s, p0/m, p1/m, z6.s, z7.s
  // 24 additional fmopa instructions
  fmopa za0.s, p0/m, p1/m, z24.s, z25.s
  fmopa za1.s, p0/m, p1/m, z26.s, z27.s
  fmopa za2.s, p0/m, p1/m, z28.s, z29.s
  fmopa za3.s, p0/m, p1/m, z30.s, z31.s
  cbnz x0, repeat_loop
// boilerplate code
mov x0, 32*512
ret
\end{lstlisting}
We see that the structure of the kernel is similar to that of the Neon kernel in Lst.\,\ref{lst:neon_fmla}.
Again, we execute a series of data processing instructions in a repeat loop.
Specifically, the loop body (lines 6-14) contains 32 FMOPA instructions, each of which performs 512 FP32 operations.
The 4096\,byte ZA array is divided into four tiles, ZA0-ZA3, each of which holds $16 \times 16$ FP32 values.
We use all four tiles repeatedly in the kernel to maximize the reuse distance.
In addition, FMOPA allows masking of the source vector registers by setting appropriate predicate registers.
The kernel uses the two predicate registers P0 and P1, and initializes all bits of both registers to 1 (lines 2 and 3) which effectively results in full outer products (no masking).

As also shown in Tab.\,\ref{tab:micro_single_core}, we observe a performance of 2009 FP32 GFLOPS when running the microbenchmark on a single performance core.
The performance drops to 502 FP32 GFLOPS when we adjust the microbenchmark to use only tile ZA0 for accumulation.
This indicates a four-cycle latency of FMOPA (non-widening) for FP32 arithmetic.

When the precision is increased to FP64, FMOPA (non-widening) computes the outer product of two 8-element vectors and adds the product to a ZA tile of $8 \times 8$ FP64 values.
Thus, each FP64 FMOPA instruction performs 128 floating-point operations.
We measured a performance of 503 FP64 GFLOPS for a corresponding microbenchmark.
In the FP64 case, it is sufficient to use four of the eight available matrix tiles for maximum performance on a single core, which again implies an instruction latency of four cycles.
If we were to lower the precision and use FP16 FMOPA (non-widening) instructions instead, we would compute the outer product of 32-element vectors.
In this case, we would have only two matrix tiles available for accumulation, but this instruction is not supported by the M4 chip.

\subsection{Scalable Matrix Extension: Matrix-Matrix Instructions}
\label{sec:micro_sme_matrix}
Another notable class of SME instructions performs matrix-matrix multiplications, where the source vector registers are interpreted as matrices.
The 8-bit integer variant of SMOPA (4-way) is such an instruction, operating on a single ZA tile, two predicate registers, and two scalable vector registers.
The instruction is widening, meaning that the datatype of the inputs has fewer bits than that of the output.
The instruction interprets the first vector register as a $16 \times 4$ matrix  and the second vector register as a $4 \times 16$ matrix of signed 8-bit integers.
The instruction multiplies the two matrices and adds the product to a matrix tile of $16 \times 16$ 32-bit integers.
This results in a total of 2048 integer operations per instruction.
Benchmarking SMOPA (4-way) on a performance core, we measured a performance of 4017 I8 GOPS which is only a 2$\times$ throughput increase over FP32.

Similarly, BFMOPA (widening) multiplies a $16 \times 2$ BF16 matrix with a $2 \times 16$ BF16 matrix and accumulates in a matrix tile with $16 \times 16$ FP32 values.
In this case, 1024 BF16 operations are performed for each instruction.
We measured a performance of 2010 BF16 GFLOPS using BFMOPA (widening), which is on par with the FP32 FMOPA (non-widening) performance.
FMOPA (widening) with FP16 inputs and FP32 output shows the same behavior as BFMOPA (widening).

\subsection{Scalable Matrix Extension: Multi-Vector Groups}
\label{sec:micro_sme_vector}
SME2 also adds the ability to execute vector instructions that operate on groups of vector registers.
As introduced in Sec.\,\ref{sec:micro_asimdneon}, an important class of vector instructions performs fused multiply-adds.
In AArch64 fused multiply-add instructions exist for Neon, SVE and SSVE under the name FMLA.
A typical FMLA instruction multiplies the values of two vector registers element-wise and adds the product to the values of a third vector register.

In contrast, SME2 supports vector instructions with two multi-vector groups, which hold up to four scalable vector registers per group.
For example, the FMLA (multiple and single vector) instruction in the FP32 variant performs four element-wise vector-vector multiplications and adds the result to a ZA single-vector group.
Thus, in FP32 arithmetic, a single instruction on M4 performs a total of $4\times32 = 128$ operations.

As can be seen in Tab.\,\ref{tab:micro_single_core}, the FP32 variant of the FMLA (multiple and single vector) instruction achieves a performance of 501 GFLOPS on a performance core, while the SSVE single-vector instruction achieves only 31 GFLOPS.
We observe the same 16$\times$ performance uplift when comparing the FP64 SME2 FMLA operating on multi-vector groups to the SSVE single-vector variant.

\subsection{Multi-Core Performance}
\label{sec:micro_multi_core} 
We have seen that a single performance core can achieve an FP32 throughput of 113\,GFLOPS when executing Neon FMLA (vector) instructions and 2009\,GFLOPS for SME FMOPA (non-widening) instructions.
This section examines the multi-core performance of M4 by running the respective microbenchmarks introduced in Sec.\,\ref{sec:micro_asimdneon} and Sec\,\ref{sec:micro_sme_outer} on 1-10 user-interactive threads.

\begin{figure} \centering{
  \includegraphics[scale=0.5]{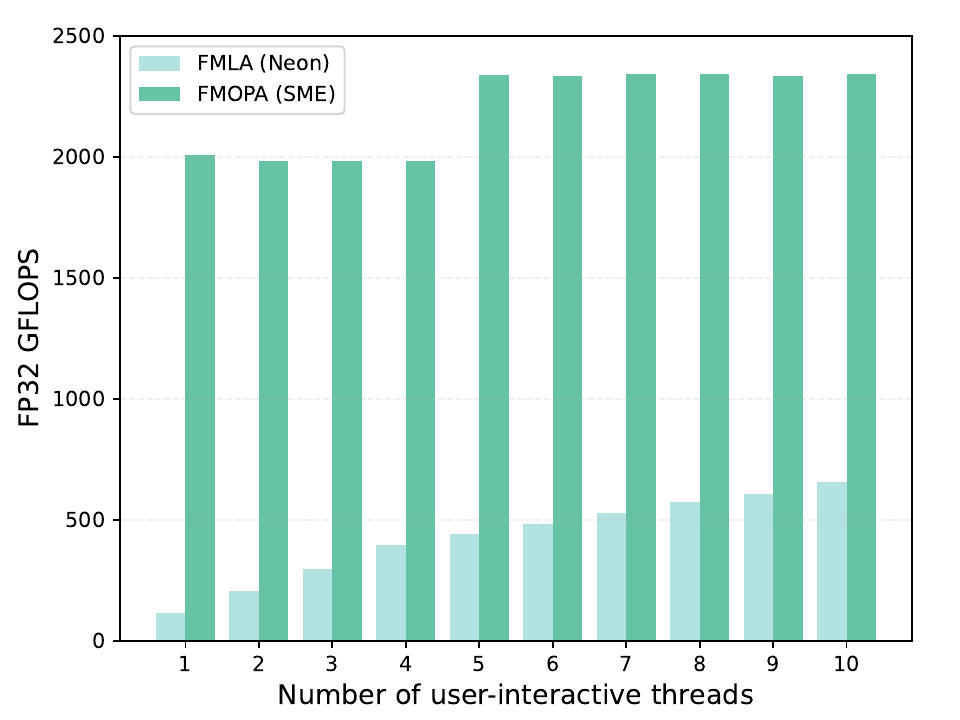}}
  \caption{Comparison of multi-core performance for the FP32 Neon FMLA (vector) instruction and the FP32 FMOPA (non-widening instruction). Performance is shown as the number of user-interactive threads increases.}
  \label{fig:micro_multicore}
\end{figure}

As shown in Fig.\,\ref{fig:micro_multicore}, we observe a Neon scaling behavior that is consistent with the four available performance and six efficiency cores.
From one to four threads, we observe an almost linear scaling with a measured performance of 395\,GFLOPS when using four threads.
After that, each new thread adds an average of 44\,GFLOPS, which is close to the standalone performance of an efficiency core (see Tab.\,\ref{tab:micro_single_core}).

The scaling behavior of the FMOPA benchmark indicates that M4 has two shared SME units.
Initially, as the number of threads increases from 1-4, the performance drops slightly from 2010 to 1983\,GFLOPS.
As soon as a fifth user-interactive thread is used, we see an increase to 2338\,GFLOPS.
The difference is close to the 357\,GFLOPS SME performance of a single efficiency core (see Tab.\,\ref{tab:micro_single_core}).
Using more than five threads does not increase SME performance further.

We ran some additional benchmarks to test our theory of two SME units in M4.
Specifically, we ran the FMOPA microbenchmark with a single user-interactive thread and a single utility thread.
However, we reduced the number of repetitions running on the utility thread to 17.8\% to match the difference observed in Tab.\,\ref{tab:micro_single_core}.
Together, the two threads achieved an FP32 performance of 2371\,GFLOPS, which is very close to the sum of the individual results in Tab.\,\ref{tab:micro_single_core}: 2009+357=2366\,GFLOPS.
Adding more threads to either thread group did not improve performance.

In summary, we measured a maximum Neon performance of 656 FP32 GFLOPS when using 10 user-interactive threads.
A single thread running SME can outperform this by up to 3.1$\times$.
With both assumed SME units, an improvement of up to 3.6$\times$ is possible.

\subsection{Bandwidth}
\label{sec:micro_bandwidth}
This section examines the bandwidth with which we can transfer data to and from the ZA array.
We measured bandwidth with simple load and store benchmarks using a single user-interactive thread.
The benchmarks load data from an FP32 array in memory to the ZA array or store data from the ZA array to memory.
We increase the size of the transferred data from 2\,KiB to 2\,GiB.
Smaller settings use the same data repeatedly, so the measured bandwidth reflects the movement of hot data without kernel startup overheads.
Note that iOS limits the amount of memory available to an application.
In our case, iOS returned about 5\,GiB of available memory to the benchmarking application.

\begin{figure} \centering{
  \includegraphics[scale=0.5]{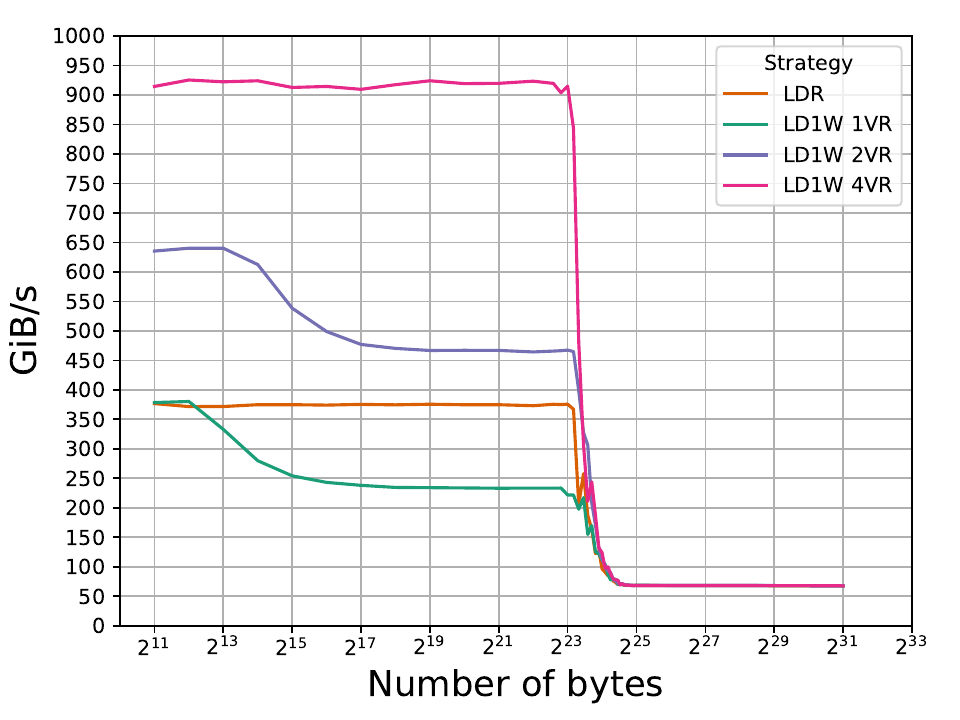}}
  \caption{Bandwidth of different strategies for loading data from memory into the ZA array. The LDR variant loads directly from memory into the ZA array, while the other strategies first load into one, two, or four vector registers (VR) and then copy the data into the ZA array. The loaded data is 128-byte aligned.}
  \label{fig:micro_bw_load}
\end{figure}
\begin{figure} \centering{
  \includegraphics[scale=0.5]{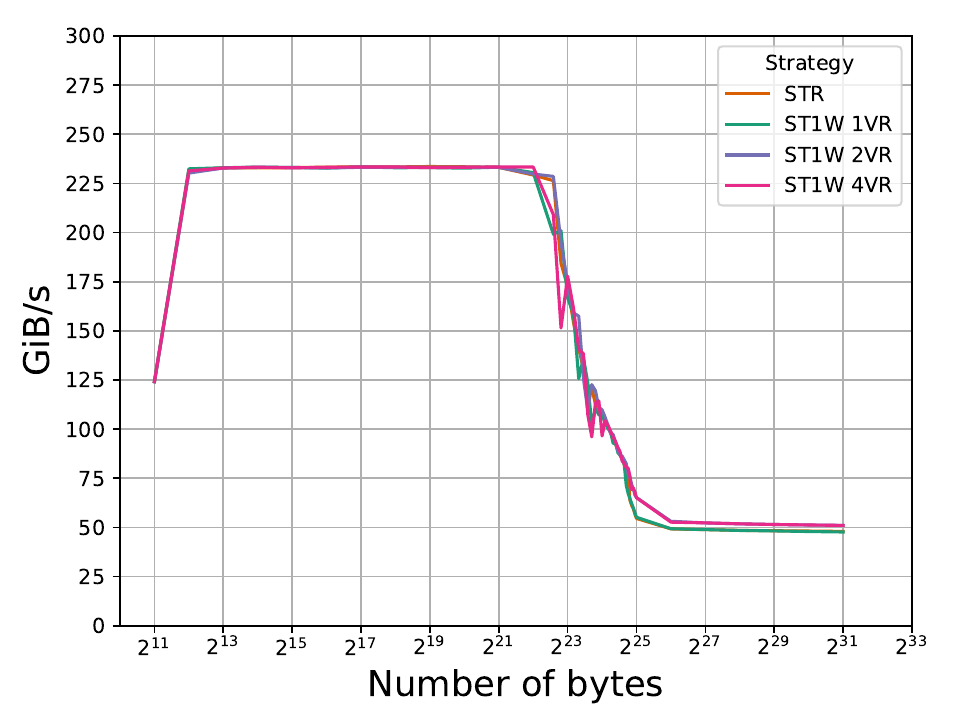}}
  \caption{Bandwidth of different strategies for storing data from the ZA array to memory. The STR variant stores directly from the ZA array to memory, while the other strategies first copy to one, two, or four vector registers (VR) and then store the data from the vector register(s) to memory. The stored data is 128-byte aligned.}
  \label{fig:micro_bw_store}
\end{figure}
Our first set of tests issues 512-bit LDR (array vector) or STR (array vector) instructions to load directly from memory into a ZA array vector, or to store data from a ZA array vector to memory.
As shown in Fig.\,\ref{fig:micro_bw_load}, we measured a bandwidth of about 375\,GiB/s for repeatedly loading up to 8\,MiB of data using the LDR (array vector) instruction.
On M4, the instruction transfers 64\,bytes from memory to the ZA array.

Fig.\,\ref{fig:micro_bw_store} shows the respective bandwidths for storing data from the ZA array to memory.
We realized direct ZA array to memory transfers using the STR (array vector) instruction, which writes 64\,bytes.
We observe the highest write bandwidth in the 4\,KiB to 4\,MiB range at about 233\,GiB/s.

\begin{lstlisting}[caption={Example code for loading data from memory into the ZA array. The code first loads the 256 bytes of data into four vector registers and then copies them into the ZA array.},label={lst:load_z_za}]
ld1w { z0.s - z3.s }, pn8/z, [x0]
mov za0h.s[w12, 0:3], { z0.s - z3.s }
\end{lstlisting}

As an alternative to using direct ZA loads, we can perform a load by first loading the data into scalable vector registers and then copying it from the vector registers into the ZA array.

An example code snippet is given in Lst.\,\ref{lst:load_z_za}. First, in line 1, the snippet loads 4$\times$64\,bytes from memory into the four consecutive registers Z0-Z3.
The memory address is stored in the general purpose register X0 and the load can be masked by the bits of the predicate register PN8.
Next, the MOV (vector to array, four registers) instruction (line 2) copies the data from the vector register to the ZA array.
Similarly, to store the data, we can first copy it from the ZA array to vector registers and then issue the appropriate store instructions on the vector registers.

Implementing this approach with indirect ZA loads gives a bandwidth of 925\,GiB/s for up to 8\,MiB of data when using 256-byte load and copy instructions (LD1W 4R).
As shown in Fig.\,\ref{fig:micro_bw_load}, the bandwidth of the 128-byte (LD1W 2VR) and 64-byte (LD1W 1VR) variants is significantly lower.
In contrast, Fig.\,\ref{fig:micro_bw_store} shows that using the indirect approach to store data from the ZA array into memory (ST1W 1VR, ST1W 2VR and ST1W 4VR) does not significantly improve bandwidth.

\begin{figure}
  \vspace{1em}
  \begin{subfigure}[htb]{0.49\columnwidth}
    \includegraphics[width=1.0\columnwidth]{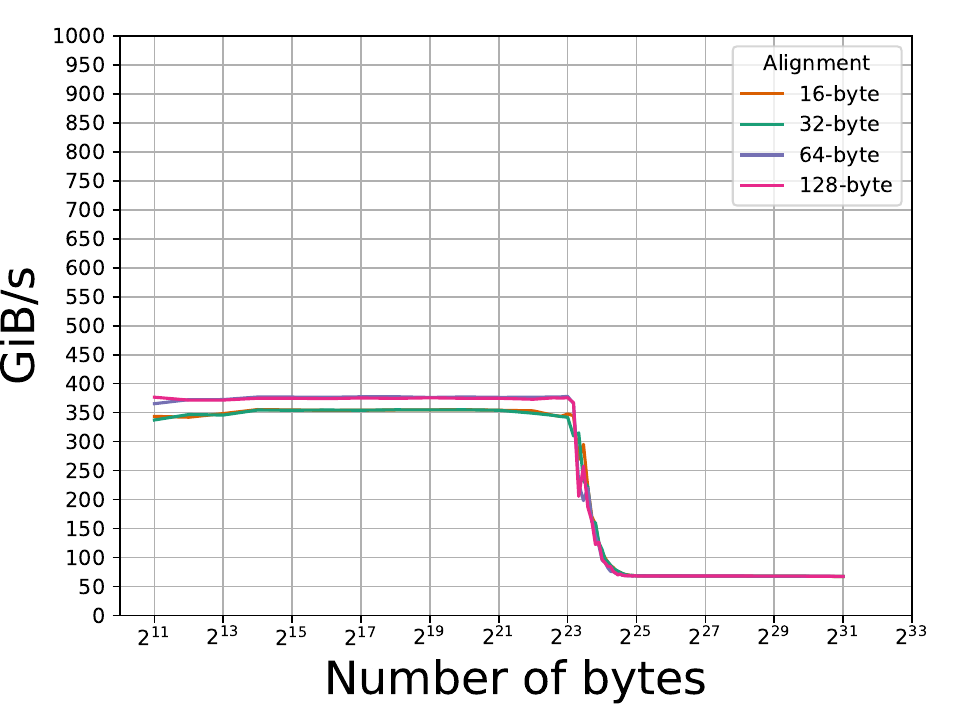}
    \caption{LDR}
  \end{subfigure}
  \begin{subfigure}[htb]{0.49\columnwidth}
    \includegraphics[width=1.0\columnwidth]{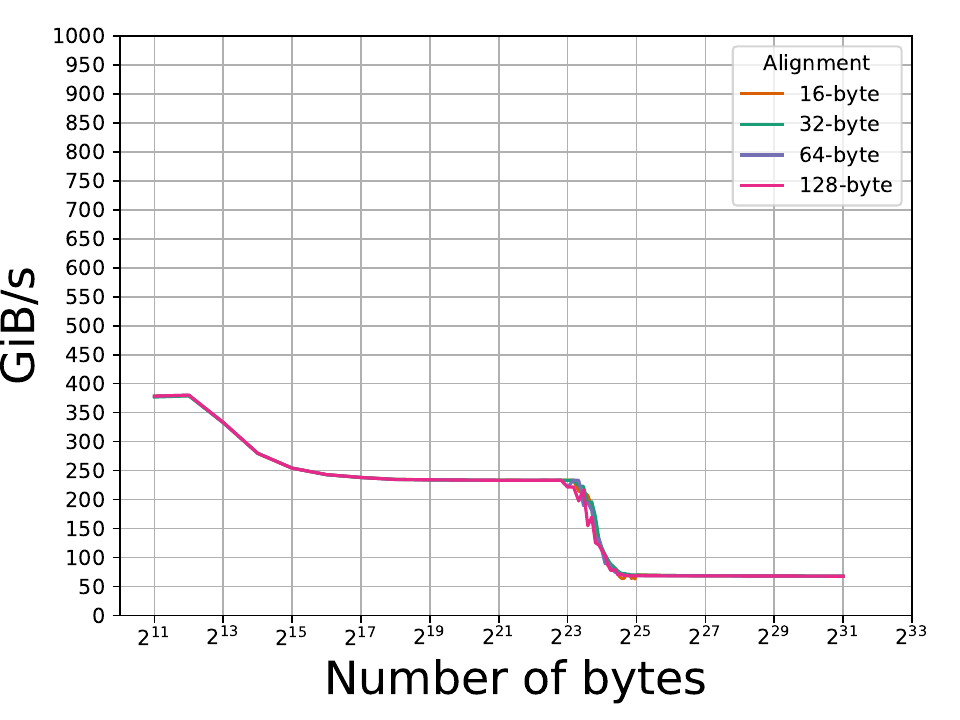}
    \caption{LD1W 1VR}
  \end{subfigure}

  \vspace{1em}
  \begin{subfigure}[htb]{0.49\columnwidth}
    \includegraphics[width=1.0\columnwidth]{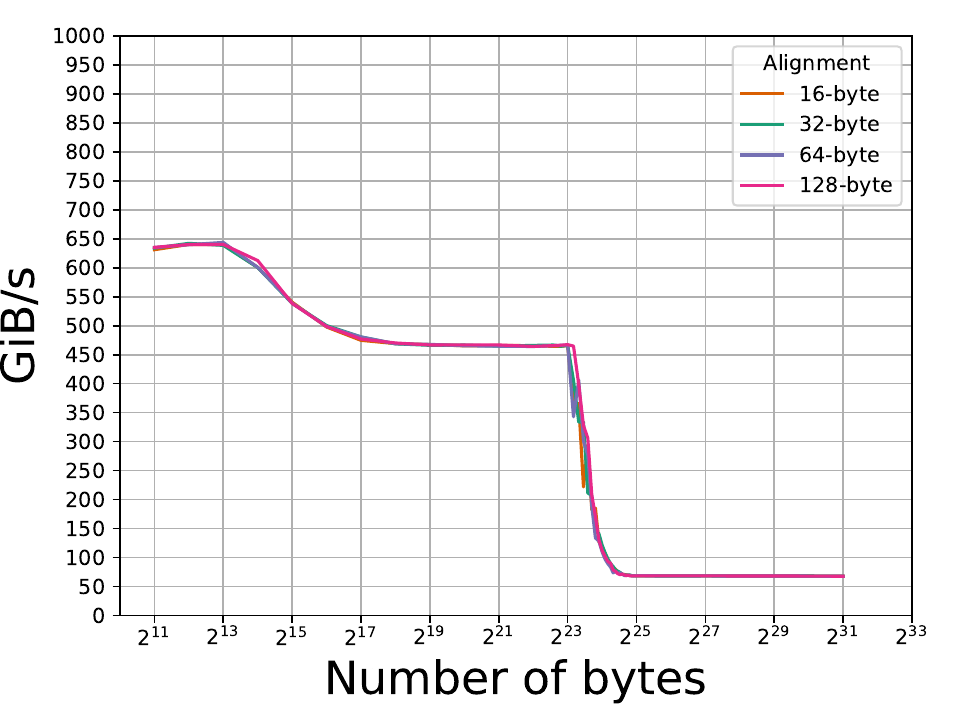}
    \caption{LD1W 2VR}
  \end{subfigure}
  \begin{subfigure}[htb]{0.49\columnwidth}
    \includegraphics[width=1.0\columnwidth]{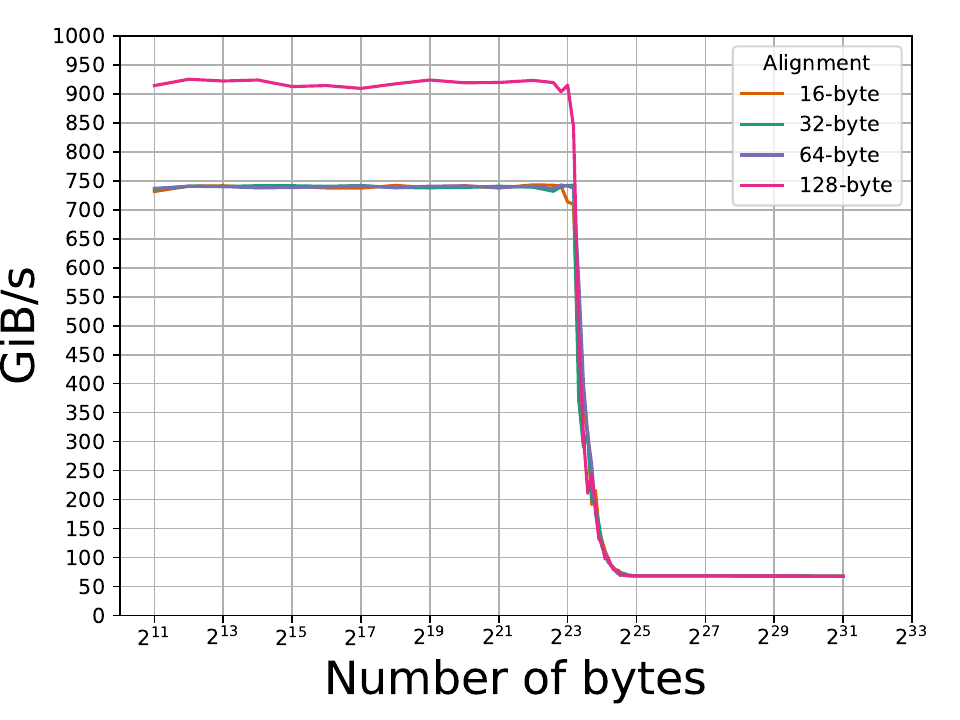}
    \caption{LD1W 4VR}
  \end{subfigure}
  \caption{Bandwidth of different strategies for loading data from memory into the ZA array, considering memory alignment. Subfigures (a) - (d) show the different load variants, where the colors denote 16-byte, 32-byte, 64-byte, and 128-byte alignment of the data.}
  \label{fig:bandwidth_load_alignment}
\end{figure}

Fig.\,\ref{fig:bandwidth_load_alignment} illustrates the studied load variants considering different memory alignments of the data.
We see that the LDR (array vector) instruction depends on the alignment.
For full read bandwidth, at least a 64-byte alignment is required.
The LD1W 4VR instruction shows a significant increase in throughput when using 128-byte alignment.
However, we do not observe any alignment impact for the indirect one-vector (LD1W 1VR) and two-vector (LD1W 2VR) variants.

\begin{figure}
  \vspace{1em}
  \begin{subfigure}[htb]{0.49\columnwidth}
    \includegraphics[width=1.0\columnwidth]{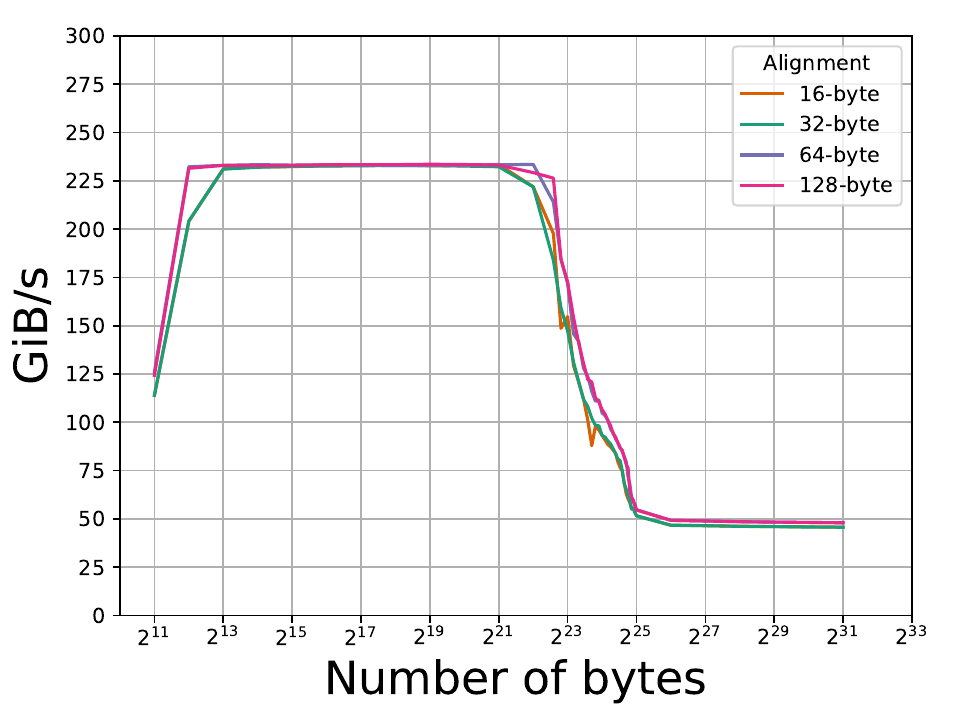}
    \caption{STR}
  \end{subfigure}
  \begin{subfigure}[htb]{0.49\columnwidth}
    \includegraphics[width=1.0\columnwidth]{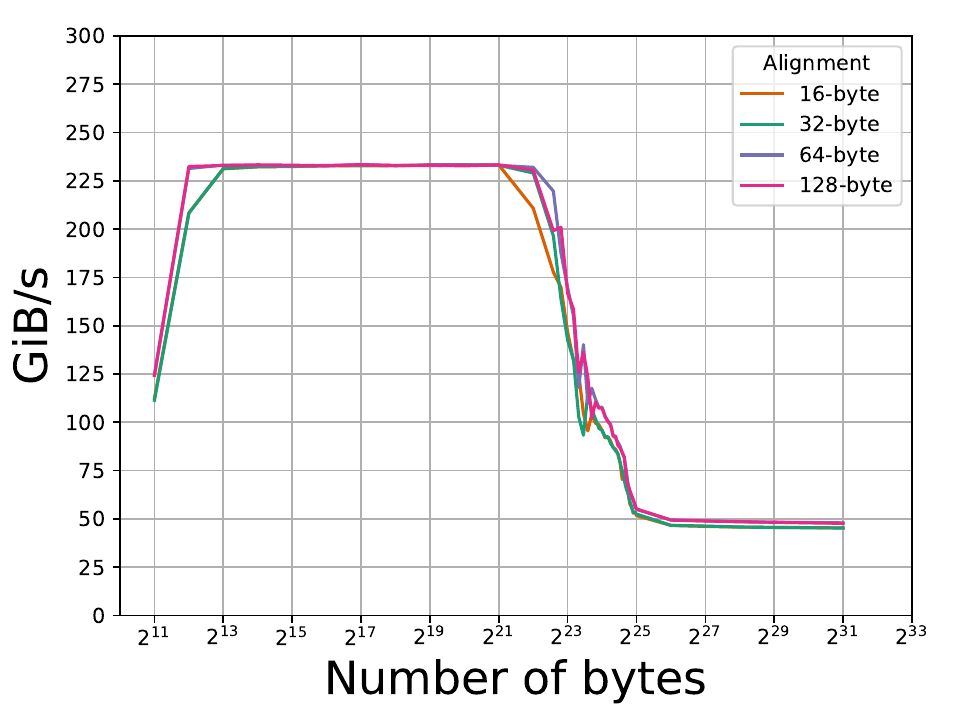}
    \caption{ST1W 1VR}
  \end{subfigure}

  \vspace{1em}
  \begin{subfigure}[htb]{0.49\columnwidth}
    \includegraphics[width=1.0\columnwidth]{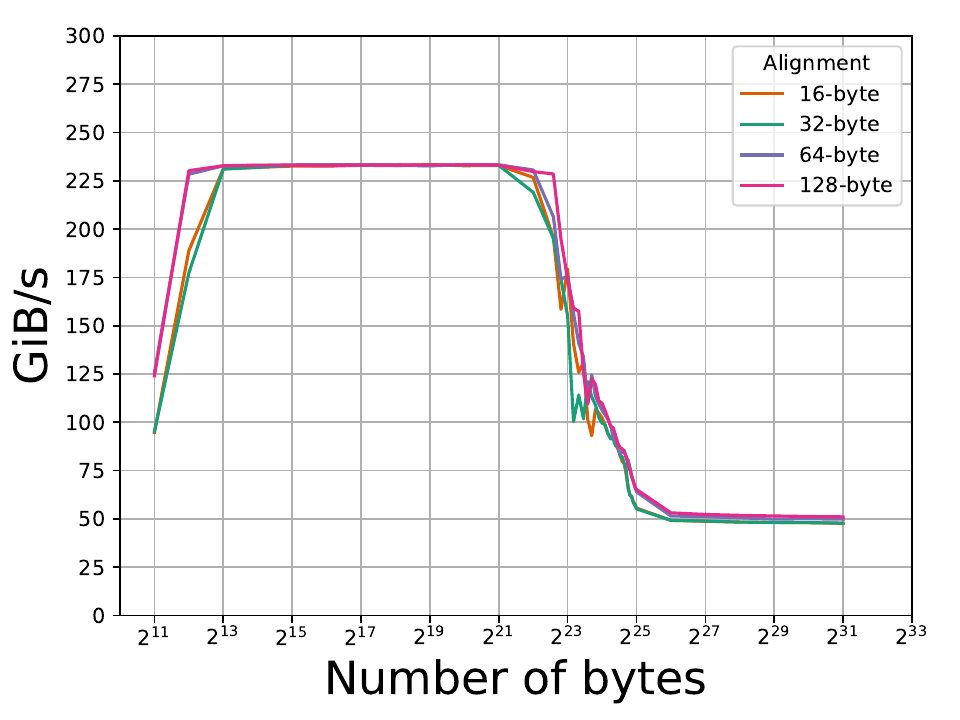}
    \caption{ST1W 2VR}
  \end{subfigure}
  \begin{subfigure}[htb]{0.49\columnwidth}
    \includegraphics[width=1.0\columnwidth]{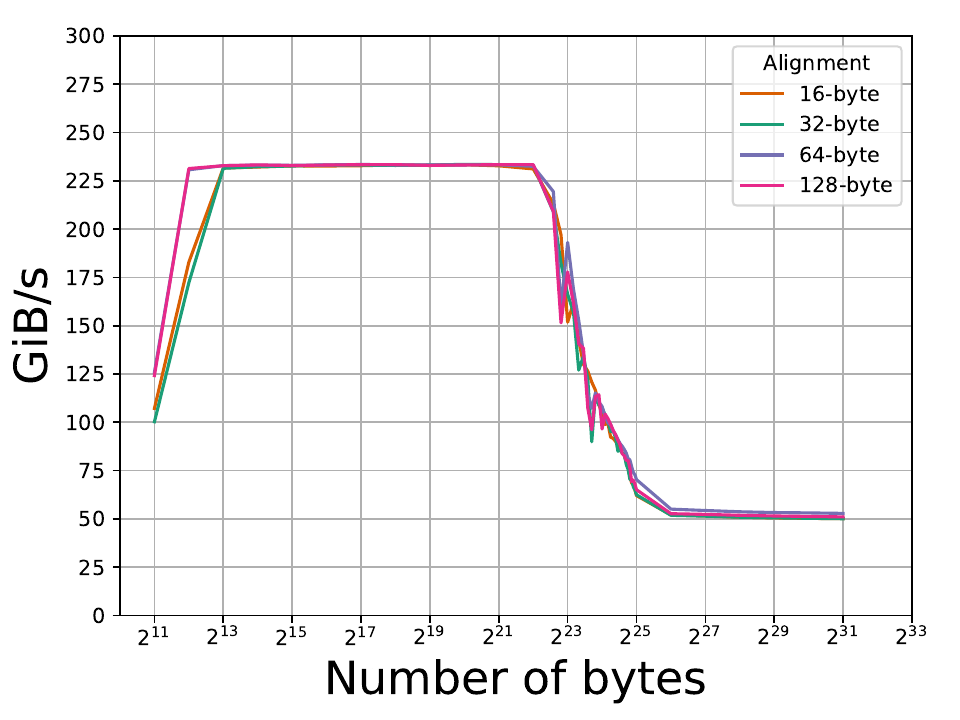}
    \caption{ST1W 4VR}
  \end{subfigure}
  \caption{Bandwidth of different strategies for storing data from the ZA array into memory, considering memory alignment. Subfigures (a) - (d) show the different store variants, where the colors denote 16-byte, 32-byte, 64-byte, and 128-byte alignment of the data.}
  \label{fig:bandwidth_store_alignment}
\end{figure}

As shown in Fig.\,\ref{fig:bandwidth_store_alignment}, alignment has a different impact on bandwidth for the store strategies studied than for the load strategies.
Most notable is an increase in bandwidth for 64-byte and 128-byte alignment when less than 8\,KiB is transferred to memory.

%% file: sections/jit.tex
\section{Just-in-time Code Generation}
\label{sec:jit}
The microbenchmarks in Sec.\,\ref{sec:micro} give us the information we need to write fast SME kernels for M4.
This section describes the extension of the LIBXSMM\footnote{LIBXSMM is available from \url{https://github.com/libxsmm/libxsmm}.} library with SME-based small GEMMs.
LIBXSMM implements a set of tensor processing primitives for all common datatypes and processors \cite{2022_libxsmm_tpp}.
Small GEMMs are among the most important primitives, and the library tailors them to the respective microarchitectures and exploits kernel metadata through just-in-time code generation.
Specifically, a LIBXSMM code generator hard-wires matrix sizes, datatypes, and leading dimensions when generating a matrix kernel.
In addition, the resulting kernel can use one of many supported SIMD extensions, such as Neon or SVE for an AArch64 processor.

We describe our LIBXSMM extension in four parts.
First, Sec.\,\ref{sec:jit_matrix_trans_b} describes the SME implementation of a GEMM that computes $C\mathrel{{+}{=}}AB^T$.
This is the simpler case, since the transposed storage of $B$ allows us to use SME's outer product instructions directly.
Second, Sec.\,\ref{sec:register_blocking} discusses our register blocking strategies.
Third, Sec.\,\ref{sec:jit_stack_trans} discusses the implementation of an SME-based GEMM $C\mathrel{{+}{=}}AB$, i.e., the transposition of $B$ is handled by the generated matrix kernel.
Finally, in Sec.\,\ref{sec:performance_eval} we evaluate the performance of our code generation approach and compare it to the BLAS implementation of the vendor-optimized Accelerate library.

\subsection{Microkernel}
\label{sec:jit_matrix_trans_b}
Motivated by the results in Sec.\,\ref{sec:micro_sme_outer} we limit the discussion in this section to FP32 FMOPA (non-widening) instructions.
\begin{figure}[htb] \centering{
  \includegraphics[scale=1.2]{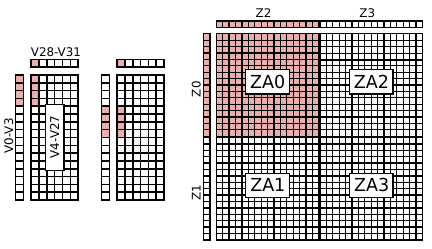}} 
  \caption{Comparison of an FP32 Neon microkernel (left) and an FP32 SME microkernel (right). The Neon kernel uses 24 128-bit accumulation registers to hold a $16\times6$ submatrix of C. The SME microkernel uses the four matrix tiles to store a $32\times32$ submatrix. Two example FMLA (by element) instructions for the Neon microkernel and a single FMOPA (non-widening) instruction for the SME kernel are highlighted in red.}
  \label{fig:fmla_fmopa}
\end{figure}
Fig.\,\ref{fig:fmla_fmopa} compares a typical Neon microkernel in LIBXSMM with an SME microkernel.
The Neon kernel uses 24 of the 32 available 128-bit vector registers to hold a $16 \times 6$ block of $C$.
In contrast, the SME microkernel uses the entire ZA array to hold an accumulator block of $32 \times 32$ elements.
Considering the required instructions, we see that the Neon code must execute an average of 64 FMLA instructions for every FMOPA instruction in the SME kernel.

\begin{lstlisting}[caption={SME microkernel that uses FMOPA (non-widerning) instructions to compute a block of GEMM $C\mathrel{{+}{=}}AB^T$.},label={lst:k_loop}]
// set predicate registers
// set register offset
k_loop:
  sub x8, x8 #0x1
  ld1w{ z0.s, z1.s }, pn8/z, [x0]
  ld1w{ z2.s, z3.s }, pn9/z, [x1]
  add x0, x0, x9
  add x1, x1, x10
  fmopa za0.s, p1/m, p0/m, z2.s, z0.s
  fmopa za1.s, p1/m, p2/m, z2.s, z1.s
  fmopa za2.s, p3/m, p0/m, z3.s, z0.s
  fmopa za3.s, p3/m, p2/m, z3.s, z1.s
  cbnz x8, k_loop
\end{lstlisting}
Assuming column-major storage for $A$ and $C$, and row-major storage for $B$, we can generate machine code similar to that in Lst.\,\ref{lst:k_loop} for an SME kernel.
The code snippet assumes that a $32 \times 32$ element block of $C$ has been loaded into the ZA array and shows the inner loop for the contraction dimension $K$.

The loop body first loads up to 32 consecutive values from a column of $A$ (line 5).
We mask the loads with the predicate register PN8 if less than 32 values are needed.
Next, we load up to 32 consecutive values from a row of $B$ (line 6).
Again, we use predication for less than 32 values.
The two instructions in lines 7 and 8 increment the addresses in X0 and X1 to point to the next column of $A$ and the next row of $B$.
The FMOPA instructions in lines 9-12 compute the four corresponding outer products and update the ZA matrix tiles.
When the K loop (lines 3, 4, and 13) is complete, we have finished computing a block of $C$.
So we write the data in the ZA array back to memory, load the next block, and start over.

\subsection{Register Blocking}
\label{sec:register_blocking}
\begin{figure} \centering{
  \includegraphics[scale=1.21]{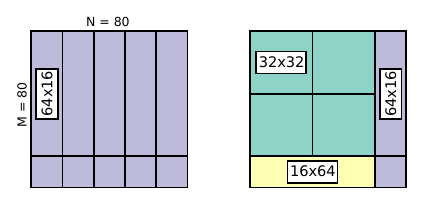}}
  \caption{Comparison of a homogeneous kernel using a single blocking strategy (left) and our heterogeneous kernel with three blocking strategies (right) for a matrix C with dimensions $M = 80$ and $N = 80$. The homogeneous kernel requires ten microkernel executions, while the heterogeneous kernel requires seven microkernel executions.}
  \label{fig:gemm_blocking}
\end{figure}
The four $16 \times 16$ element ZA matrix tiles can be used for different register blocking strategies.
In Sec.\,\ref{sec:jit_matrix_trans_b} we discussed a microkernel that uses a $32 \times $32 element blocking of the output matrix.
The advantage of this strategy is that it only needs to load 32 column values of $A$ into two Z registers and 32 row values of $B$ into two Z registers for a complete update of the accumulator block.
This means that the kernel loads a total of $32+32=64$ values of $A$ and $B$ for each update of the ZA matrix array.
This strategy is also shown in Fig.\,\ref{fig:fmla_fmopa}, where Z0 and Z1 are used for the values of $A$ and Z2 and Z3 for those of $B$.

Our code generator supports two additional blocking schemes.
Specifically, the generator supports $16 \times 64$ register blocking and $64 \times 16$ register blocking.
Both blocking schemes require loading $16+64=64+16=80$ values of $A$ and $B$ for a complete accumulator block update. 
However, the advantage of having different blocking schemes is that we can choose and mix them according to the sizes of the matrices in the small GEMM.
For example, assume $M=80$, $N=80$, i.e. $C \in \mathbb{R}^{80 \times 80}$.
In this case, we generate code that uses all three of our blocking strategies.

The resulting heterogeneous kernel performs seven microkernel executions to compute the corresponding small GEMM.
As shown on the right side of Fig.\,\ref{fig:gemm_blocking}, the first 64 columns of $C$ are computed by four executions of a microkernel with $32 \times 32$ register blocking and one execution of a microkernel that uses $16 \times 64$ register blocking.
The sixth kernel execution uses a $64 \times 16$ register blocking to compute the first 64 rows of the last 16 columns of $C$.
Finally, the seventh microkernel execution computes the remaining $16 \times 16$ values of the matrix $C$.
Here we use a heavily masked $64 \times 16$ blocking.

\subsection{Transposing B}
\label{sec:jit_stack_trans}
The matrix kernels described in Sec.\,\ref{sec:jit_matrix_trans_b} and Sec.\,\ref{sec:register_blocking} compute GEMMs of the form $C\mathrel{{+}{=}}AB^T$.
This means that $A$ and $C$ are assumed to be column-major, while $B$ is assumed to be row-major.
In line 6 of Lst.\,\ref{lst:k_loop}, the row-major storage of $B$ allows us to simply load up to 32 consecutive row values of $B$.
These values can then be used directly for the outer products in lines 9 -12.

For a column-major $B$, we have to solve the problem that two consecutive row values have a stride determined by the leading dimension of $B$.
We have not found an elegant way to perform efficient transpositions using only registers.
Instead, we allocate scratch memory on the stack and use it to store a transposed panel of $B$.
A similar strategy is used in the SME Programmer's Guide \cite{2024_sme_guide}.

Assuming that $K$ is the number of rows in $B$ and that our kernel operates on entire $32 \times 32$ blocks of elements, we transpose a $K \times $32 panel of B.
To do this, we block the panel into $16 \times 16$ blocks.
We copy each block by horizontal MOV (tile to vector, four registers) instructions from the vector registers to a ZA tile and back by vertical MOV (vector to tile, four registers) instructions.
\begin{lstlisting}[caption={Assembly kernel transposing a $16 \times $16 block using the ZA array. First, the data is copied from vector registers Z0-Z15 to tile ZA0 using the horizontal view. Then the data is copied back to the same vector registers using the vertical view.},label={lst:trans}]
mov w12, #0
mov za0h.s[w12, 0:3], {z0.s - z3.s}
add w12, w12, #4
mov za0h.s[w12, 0:3], {z4.s - z7.s}
add w12, w12, #4
mov za0h.s[w12, 0:3], {z8.s - z11.s}
add w12, w12, #4
mov za0h.s[w12, 0:3], {z12.s - z15.s}
mov w12, #0
mov { z0.s - z3.s },   za0v.s[w12, 0:3]
add w12, w12, #4
mov { z4.s - z7.s },   za0v.s[w12, 0:3]
add w12, w12, #4
mov { z8.s - z11.s },  za0v.s[w12, 0:3]
add w12, w12, #4
mov { z12.s - z15.s }, za0v.s[w12, 0:3]
\end{lstlisting}
Lst.\,\ref{lst:trans} shows a code snippet that transposes a $16 \times 16$ block.
Lines 1-8 copy a column-major block from vector registers Z0-Z15 to ZA0 using the horizontal view.
Next, we copy the data in the matrix tile ZA0 back to Z0-Z15, but using the vertical view (lines 9-16).
In effect, using both the horizontal and vertical copy transposes the block, which we then store in our scratch memory.

We align the blocks in the scratch memory to 64-byte boundaries.
Once an entire $32 \times K$ panel of $B$ is transposed, we multiply it by $A$ using the procedure described in Sec.\,\ref{sec:jit_matrix_trans_b}.

\subsection{Performance Evaluation}
\label{sec:performance_eval}
\begin{figure} \centering{
  \includegraphics[scale=0.5]{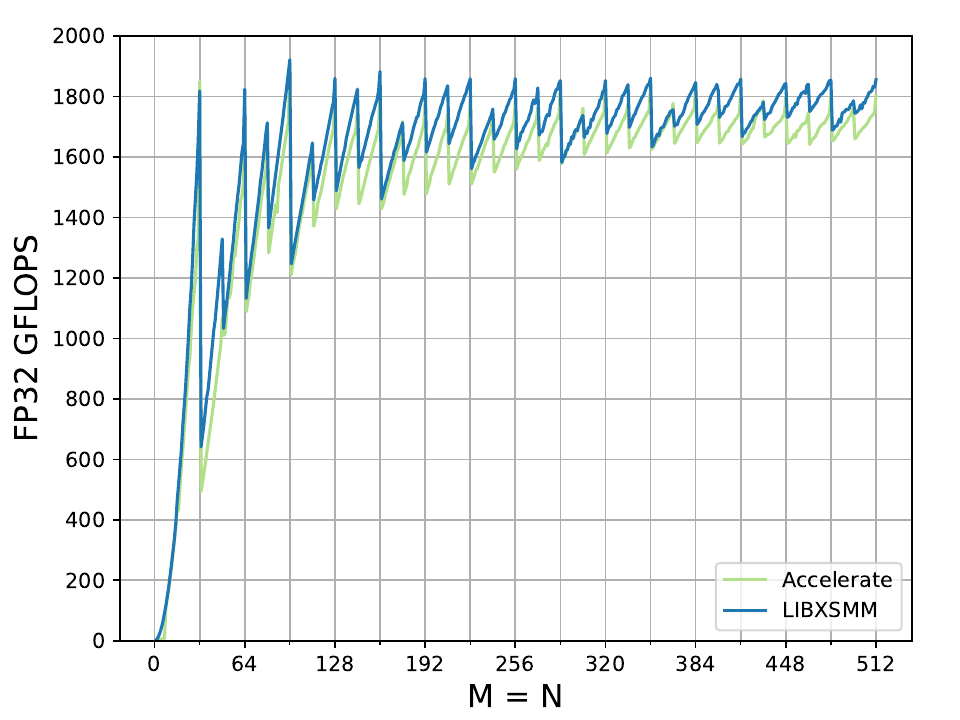}}
  \caption{FP32 performance comparison of our code generator with Accelerate BLAS for small GEMMs. The sizes of the output matrices are given on the x-axis (M=N). The contraction dimension K has a fixed size of K=512. Matrices A and C are assumed to be column-major and matrix B is assumed to be row-major.}
  \label{fig:jit_gemm_b_trans}
\end{figure}
Fig.\,\ref{fig:jit_gemm_b_trans} compares the performance of our generated kernels with the SGEMM implementation in Accelerate BLAS.
Shown are the GEMMs $C\mathrel{{+}{=}}AB^T$ with $M, N \in [1, 2, \ldots, 512]$ and $K=512$.
We see that our approach outperforms the vendor library in almost all tested settings.

\begin{figure} \centering{
  \includegraphics[scale=0.5]{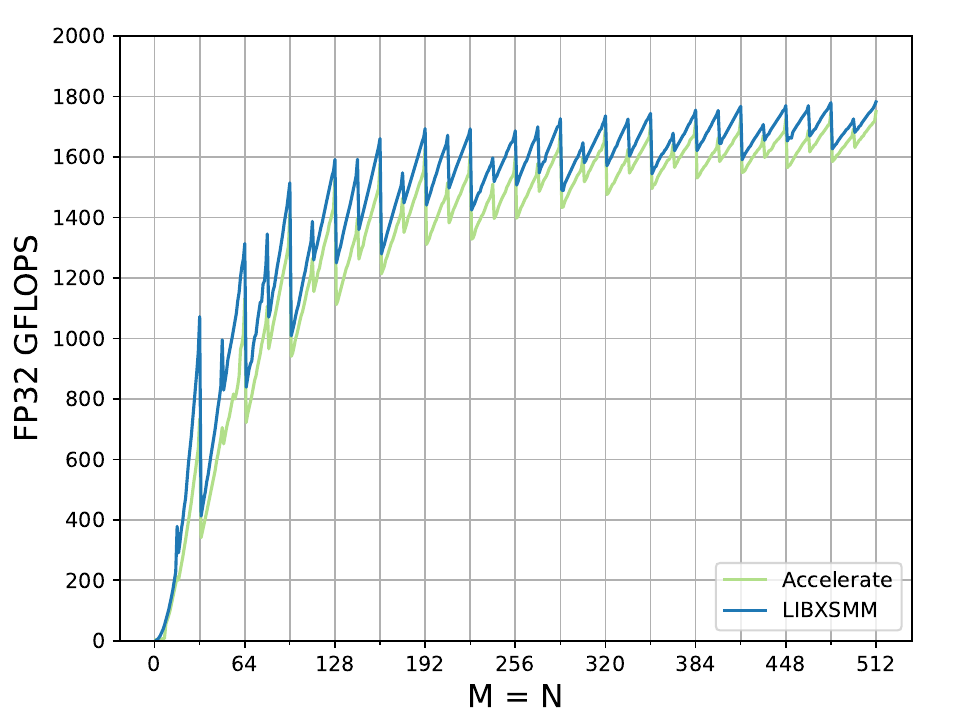}} 
  \caption{FP32 performance comparison of our code generator with Accelerate BLAS for small GEMMs. The sizes of the output matrices are given on the x-axis (M=N). The contraction dimension K has a fixed size of K=512. All matrices are assumed to be column-major.}
\label{fig:jit_gemm_col_maj}
\end{figure}
Fig.\,\ref{fig:jit_gemm_col_maj} shows the performance of the same settings for a column-major matrix $B$, i.e. the GEMM $C\mathrel{{+}{=}}AB$ with $M, N \in [1, 2, \ldots, 512]$ and $K=512$.
As described in Sec.\,\ref{sec:jit_stack_trans}, the kernel must now transpose the matrix $B$ in order to use outer product FMOPA instructions.
In this case, our implementation outperforms Accelerate BLAS in all tested settings.

%% file: sections/outlook.tex
\section{Discussion and Outlook}
\label{sec:outlook}
We have presented a thorough performance analysis of M4's CPU, in particular the matrix acceleration enabled by the Scalable Matrix Extension (SME) of the Arm Architecture.
The microbenchmarks conducted allow us to outline the strengths and weaknesses of the SME implementation in M4, and guided the design of a code generator for fast SME-enabled tensor processing primitives.

In FP32 arithmetic, throughput of over 2.3\,TFLOPS can be achieved using outer product SME instructions.
This is a 3.6$\times$ improvement over the multi-core Neon performance of the CPU.
The SME performance of other datatypes (FP64, FP16, BF16, I8) on M4 is comparatively low.
Only for I8 inputs a moderate twofold speedup is possible, making M4's SME acceleration FP32-centric.

Our bandwidth benchmarks show that it is advantageous to first load data into the scalable vector registers and then copy it into the matrix array, instead of loading directly into the matrix array.
Using this two-step method we obtained a 2.6$\times$ improvement in read bandwidth over direct loads from the L2 cache.

The results of the benchmarks performed guided our extension of the LIBXSMM library with a code generator for SME-based small GEMMs.
By benchmarking our code generator for a variety of matrix sizes, we were able to demonstrate competitive performance results.
In the case where $A$ and $C$ are stored in column-major format and $B$ in row-major format, our implementation is faster than the vendor-optimized BLAS routines in Accelerate in almost all tested settings.
When all matrices are stored in column-major format, our implementation outperforms Accelerate BLAS in all tested settings.

Apple M4 is the first chip to support SME.
This is not surprising since the design of SME is similar to Apple AMX.
We expect to see SME support in all upcoming Apple silicon, especially in upcoming M-series SoCs.
Higher throughput of reduced-precision SME instructions (FP16, BF16, I8) could further accelerate CPU-native machine learning inference.
At the same time, we expect other vendors to support SME in the near future.
For example, the benefits of SME become apparent when considering that Nvidia's 72-core Grace CPU has only 3.5$\times$ higher FP32 performance than M4's SME unit associated with the performance cores.
In particular, low-latency HPC workloads that require dense linear algebra would benefit greatly from matrix acceleration.